\documentclass[showpacs,amsmath,amssymb,aps,showkeys,floatfix,prd,a4paper]{revtex4}
\usepackage{subfigure}
\usepackage{dcolumn}
\usepackage{bm}
\usepackage{epsfig}
\usepackage{amsfonts}
\usepackage{amssymb,amscd}
\usepackage{multirow}

\usepackage{xcolor}
\usepackage[normalem]{ulem} 

\newcommand{\be}{\begin{equation}}
\newcommand{\ee}{\end{equation}}
\def\beq{\begin{equation}}
\def\eeq{\end{equation}}
\def\beqa{\begin{eqnarray}}
\def\eeqa{\end{eqnarray}}
\newcommand{\ba}{\begin{eqnarray}}

\begin{document}




\title{Production of $\eta_b$ in ultra-peripheral $PbPb$ collisions}

\author{C.N. Azevedo$^{1}$, F.C. Sobrinho$^{1}$, F.S. Navarra$^{1}$\\
$^1$Instituto de F\'{\i}sica, Universidade de S\~{a}o Paulo,
Rua do Mat\~ao 1371 - CEP 05508-090,
Cidade Universit\'aria, S\~{a}o Paulo, SP, Brazil\\
}


\begin{abstract}

Very recently, the two-photon decay width of the  $\eta_b$ meson was computed with lattice QCD methods. This decay has not yet been measured. 
The knowledge of this width allows for the calculation of the  $\eta_b$ production cross section through photon-photon interactions in ultraperipheral $PbPb$ collisions. In this work  we present this calculation, which is the first application of the lattice result. Since 
UPCs are gaining an increasing attention of the heavy ion community, we take the opportunity to perform a comprehensive study of the different 
ways of defining ultra-peripheral collisions and of the different ways to treat the  equivalent photon flux.  

\end{abstract}



\maketitle

\vspace{0.5cm}


\section{Introduction}

Ultra-peripheral heavy ion collisions (UPHICs) provide an opportunity to improve our understanding of the Standard Model as well as to search 
for New Physics \cite{bertulani:1987tz, bertulani:2005ru, goncalves2005sn, baltz:2007kq, contreras2015dqa, klein2020fmr}. 
%
In these collisions the incident nuclei do not overlap, which implies the suppression of the strong interactions and the dominance of the electromagnetic interaction between them. 
Over the last years, the study of photon induced processes in hadronic colliders has become a reality with a large amount of experimental results published for different final states. New states are expected to be seen in the future. 
The essential  feature of these processes is that relativistic heavy ions give rise to strong electromagnetic fields, so that in a hadron-hadron collision, photon-hadron and photon-photon interactions can occur and they  may lead to the production of particles.
Moreover, processes involving photons can be exclusive, where the resulting final state is very clean. 
A typical example of exclusive process is the production of pseudoscalar mesons due to two photon fusion.  The resulting final state is very simple, consisting of a pseudoscalar meson with very small transverse momentum, two intact nuclei, and two rapidity gaps, i.e., empty regions in pseudorapidity that separate the intact very forward nuclei from the produced state. Such aspects can, in principle, be used to separate the events and to test predictions.

Very recently, the decay  $\eta_b \rightarrow \gamma\gamma$  was studied in lattice QCD for the first time in Ref. \cite{Colquhoun:2024wsj}, and the decay width  was calculated, providing an accurate prediction to be tested at Belle II. 
As a first application of this result, the first purpose of this work is to compute the $\eta_b$ production cross section through
$\gamma\gamma$ interactions in ultra-peripheral $PbPb$ collisions at the LHC energy $\sqrt{s} = 5.02$ TeV.

A second purpose of our work is to perform a comprehensive comparison of the ingredients used in this type of calculation: the practical  
definition of UPC, the method  applied to obtain the equivalent photon flux and the form factor of the photon source.

\section{Formalism}

\begin{figure}[t]
    \centering
    \includegraphics[width=.45\linewidth]{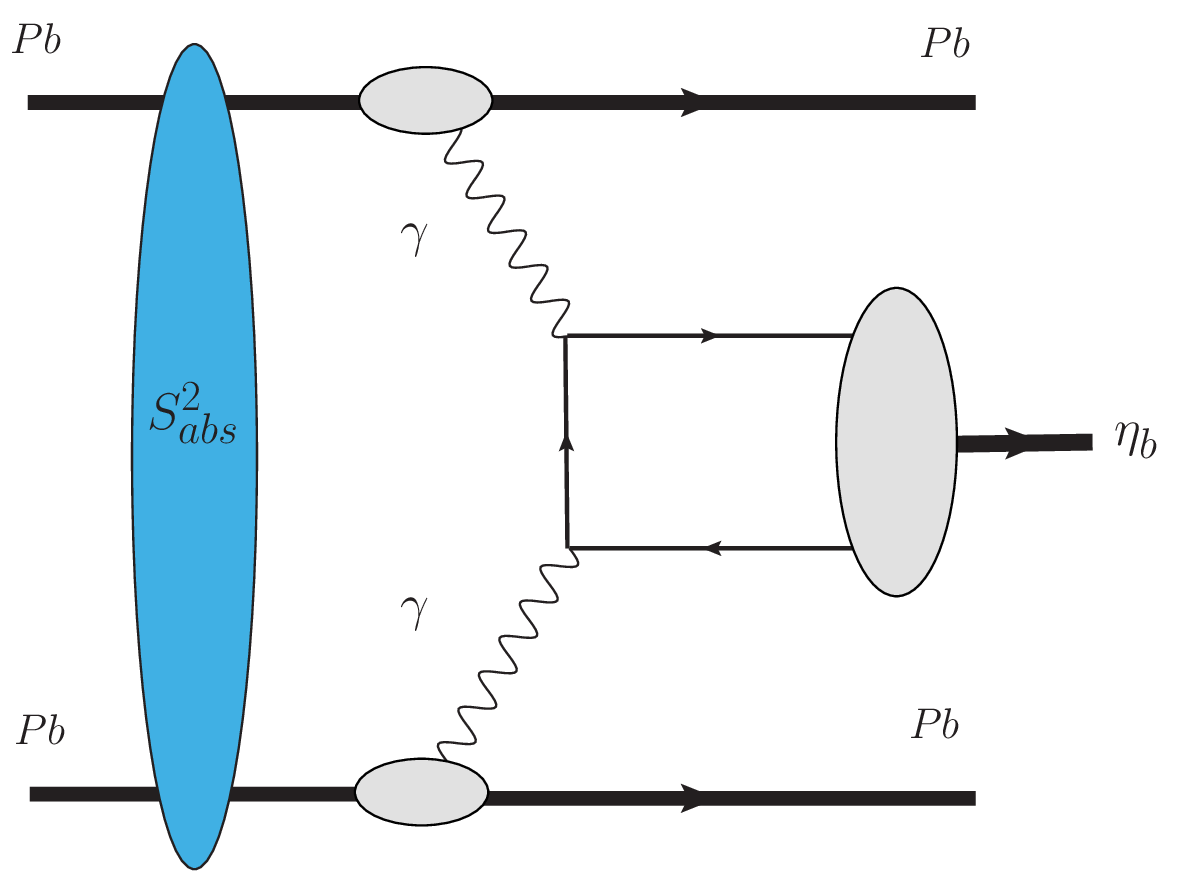}
    \caption{Production of $\eta_b$ by $\gamma\gamma$ interaction in ultra-peripheral $Pb Pb$ collisions.}
    \label{fig:photon-photon_etab}
\end{figure}

In an UPC, the intense electromagnetic fields that accompany the relativistic heavy ions can be viewed as a spectrum of equivalent photons and $\eta_b$ can be produced through the $\gamma\gamma \rightarrow \eta_b$ process (see Fig. \ref{fig:photon-photon_etab}). The photon flux is proportional to the square of the nuclear charge $Z$ and the associated cross section to $Z^4$, implying large cross sections at LHC energies.

\subsection{The production cross section}

\begin{figure}[t]
    \centering
    \includegraphics[width=.45\linewidth]{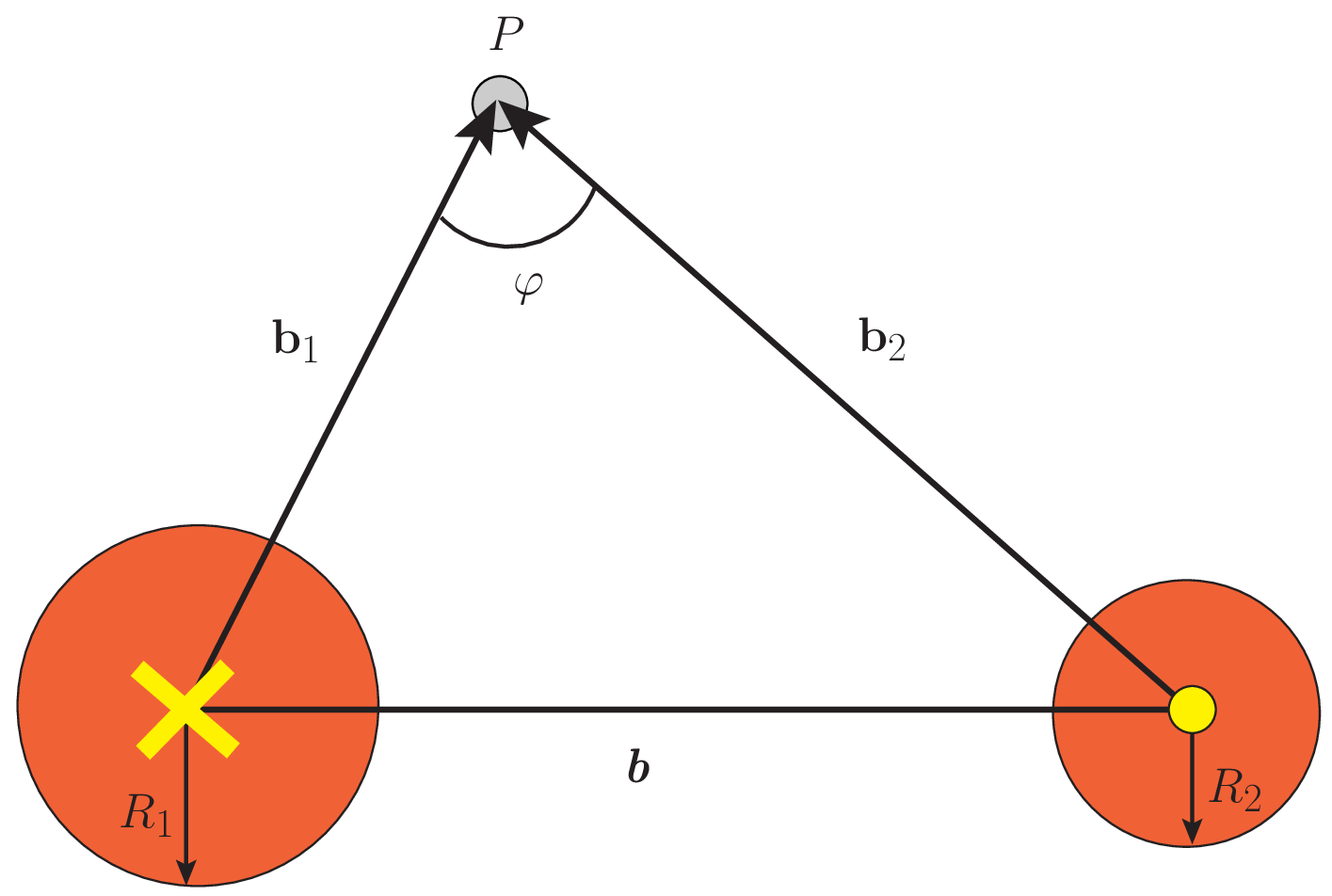}
    \caption{
    View in the plane perpendicular to the direction of motion of the two ions. In a semiclassical picture the two equivalent photons of energy $\omega_1$ and $\omega_2$ collide in a point $P$ with distance $\mathbf{b}_1$, from ion 1 and $\mathbf{b}_2$ from ion 2. The impact parameter $\textit{\textbf{b}}$ is the distance between the colliding nuclei with radius $R_1$ and $R_2$.
    } 
    \label{fig:parameter_b} 
    \end{figure}

Initially, let us present a brief review of the formalism needed to describe the pseudoscalar meson production in $\gamma\gamma$ interactions in ultraperipheral $PbPb$ collisions. 
In the equivalent photon approximation, the total cross section for the production of $\eta_b$ 
can be factorized in terms of the equivalent photon flux of each of the nuclei and the photoproduction cross section, as follows: 
\begin{equation}
    \sigma(PbPb \rightarrow Pb \otimes \eta_b \otimes Pb; s) = 
    \int 
    \hat{\sigma}(\gamma\gamma \rightarrow \eta_b; W) 
    \, 
    N(\omega_1,\mathbf{b}_1)  \, N(\omega_2,\mathbf{b}_2) \, 
     S_{abs}^2(\mathbf{b})  \, \mbox{d}^2 \mathbf{b}_1  \, \mbox{d}^2 \mathbf{b}_2  \, \mbox{d}\omega_1  \, \mbox{d}\omega_2
    \, ,
\label{eq:total_cross_section_etab}
\end{equation}
where $\sqrt{s}$ is the center-of-mass energy for the collision $PbPb$, $\otimes$ characterizes a rapidity gap in the final state, $W= \sqrt{4\omega_1\omega_2}$ is the invariant mass of the $\gamma\gamma$ system, and $\hat{\sigma}(\gamma\gamma\rightarrow \eta_b)$ is the photoproduction cross section of the $\eta_b$ due to the fusion of two photons. Moreover, $\omega_i$ is the energy of the photon emitted by nucleus $A_i$ at an impact parameter, or distance, $b_i$ from $A_i$. The photons interact at the $P$ point shown in Fig. \ref{fig:parameter_b}. 
Note that the photon energies $\omega_1$ and $\omega_2$ are related to $W$ and the rapidity $Y$ of the produced state, through  
\begin{equation}
    \omega_1 = 
    \frac{W}{2} e^Y 
    \qquad \mbox{and} \qquad 
    \omega_2 = 
    \frac{W}{2}e^{-Y}
    \,. 
\label{eq:omega}
\end{equation}
As a consequence,  the total cross section can be rewritten as (for details see e.g. Ref. \cite{Antoni:2010})
\begin{equation}
    \sigma(PbPb \rightarrow Pb \otimes \eta_b \otimes Pb; s) = 
    \int 
    \hat{\sigma}(\gamma\gamma \rightarrow \eta_b; W) 
    \, 
    N(\omega_1,\mathbf{b}_1) \, N(\omega_2,\mathbf{b}_2) \, 
     S_{abs}^2(\mathbf{b}) \frac{W}{2} \, 
    \mbox{d}^2 \mathbf{b}_1  \, \mbox{d}^2 \mathbf{b}_2  \, \mbox{d}W \, \mbox{d}Y
    \, .
\label{eq:cross_section_WY}
\end{equation}
Using the Low formula \cite{Low:1960wv}, we can express the cross section for the photon-photon interaction producing the pseudoscalar meson in terms of the two-photon decay width as follows: 
\begin{equation}
    \hat{\sigma}_{\gamma\gamma\rightarrow \eta_b}(\omega_1,\omega_2) = 
    8\pi^2 (2J + 1) \,
    \frac{\Gamma_{\eta_b \rightarrow \gamma\gamma}}{M_{\eta_b}} \,
    \delta(4\omega_1\omega_2 - M_{\eta_b}^2) 
    \,,
\label{eq:low_etab}
\end{equation}
where $M_{\eta_b}$ and $J$ are  the mass and spin of the produced $\eta_b$ respectively.
The decay rate for $\eta_b \rightarrow \gamma\gamma$ was calculated in Ref. \cite{Colquhoun:2024wsj} using lattice QCD and  was 
found to be  $\Gamma(\eta_b \rightarrow \gamma\gamma) = 0.557 (32)(1)$ keV.

\subsection{The photon flux} 

The equivalent photon flux $N(\omega_i,b_i)$ of photons with energy $\omega_i$ at a transverse distance $b_i$ from the center of the nucleus, defined in the plane transverse to the trajectory, can be expressed in terms of the charge form factor $F(q)$, where 
 $q$ is the four-momentum of the quasireal photon. It reads: 
\begin{equation}
N(\omega, b) =  \frac{Z^2 \, \alpha}{\pi^2 \, \omega \,  b^2} 
\left[ \int_{0}^{\infty} u^2 \,  J_1(u) \, F \left( \sqrt{\frac{u^2 + (b \, \omega / \gamma)^2}{b^2} } \, \right)
                        \frac{1}{u^2 + (b \, \omega / \gamma)^2 } \; du
\right]^2 
\label{eq:flux}
\end{equation}
where  $\alpha= e^2/(4 \, \pi)$, $J_1$ is the Bessel function of the first kind and $\gamma$ is the Lorentz factor of the photon source 
($\gamma = \sqrt{s}/2m_p$ and $m_p$ is the proton mass). 
In the case of a nucleus-nucleus collision, the realistic form factor is obtained 
as a Fourier transform of the nuclear charge density, and is analytically expressed by: 
\begin{equation}
F(q^2) = \frac{4 \, \pi}{A \, q^3} \rho_0 
[\sin(q \, R) - q \, R \, \cos(q \, R)] 
\left[
    \frac{1}{1+q^2 \,  a^2}
\right] 
\,,
\label{eq:realistic}
\end{equation}
with the parameters $a = 0.549$ fm and $\rho_0 = 0.1604/A$ $\text{fm}^{-3}$ obtained for the lead nucleus \cite{DeJager:1974liz, Bertulani:2001zk}. A simpler form factor, often used in the literature, is of the monopole type, given by 
\begin{equation}
    F(q^2) = 
    \frac{\Lambda^2}{\Lambda^2 + q^2}
    \, , 
    \qquad \mbox{with} \qquad \Lambda_{Pb} = 0.088 \, \mbox{GeV}
\label{eq:monopole}
\end{equation}
where $\Lambda$ is a constant adjusted to reproduce the root-mean-square (rms) radius of a nucleus \cite{Antoni:2010}. Introducing the monopole form factor into (\ref{eq:flux}), we obtain the following expression for the photon flux
\begin{equation}
    N(\omega, b) = 
    \frac{Z^2\alpha}{\pi^2\omega} 
    \left[
        \frac{\omega}{\gamma} \, K_1 
        \left(
            \frac{b \omega}{\gamma}
        \right) 
        - \sqrt{
        \left( 
            \frac{\omega^2}{\gamma^2} + \Lambda^2
        \right)}
        K_1
        \left(
            b \, 
            \sqrt{\frac{\omega^2}{\gamma^2} + \Lambda^2}
        \right)
    \right]^2
    \, .
\label{eq:monopole_flux}
\end{equation}
On the other hand, assuming a point-like form factor ($F = 1$), the photon flux takes the following form 
\begin{equation}
    N(\omega,b) = 
    \frac{Z^2\alpha}{\pi^2 \omega b^2} 
    \left( 
        \frac{\omega b}{\gamma}
    \right)^2
    \left[
        K_1^2
        \left(
            \frac{\omega b}{\gamma}
        \right)
        + \frac{1}{\gamma^2} K_0^2 
        \left( 
            \frac{\omega b}{\gamma}
        \right)
    \right]
    \, ,
\label{eq:pointlike_flux}
\end{equation}
where $K_0$ and $K_1$ are the modified Bessel functions. It is important to emphasize that this flux diverges at $b \rightarrow 0$. 
In this case, we need to take a lower limit cut for the integrals over $b$. Usually, the integration is performed from a minimum distance 
$b_{min} = R$ \cite{Bertulani:2009qj}. As demonstrated in Ref. \cite{Antoni:2010}, the realistic and monopole form factors are similar 
within a limited range of $q$ and differ at large $q$. Additionally, the point-like form factor is an unrealistic approximation, as it 
disregards the internal structure of the nucleus. 

\subsection{On the practical definition of  an ultra-peripheral collision}

\subsubsection{Pure geometry}

The factor $S^2_{abs}(b)$ in (\ref{eq:total_cross_section_etab}) depends on the impact parameter of the collision and is denoted the absorptive factor, which excludes the overlap between the colliding nuclei and selects only ultra-peripheral collisions. A widely used procedure to exclude the strong interactions between the incident nuclei was proposed by Baur and Ferreira-Filho \cite{baur1990}. They assumed that: 
\begin{equation}
    S_{abs}^2(\mathbf{b}) = 
    \Theta (|\mathbf{b}| - R_1 - R_2) = 
    \Theta (|\mathbf{b}_1 - \mathbf{b}_2| - R_1 - R_2)
    \, ,
\label{eq:Sabs_baur}
\end{equation}
where $R_i$ is the nuclear radius. In this model, the probability to have a hadronic interaction when $b > R_1 + R_2$ is zero. This 
procedure, while intuitive, introduces a sharp cut in the calculation and tends to overemphasize the role of geometry. It is a based 
on a probably too classical picture of nuclear collisions. In addition, it does not contain any energy dependence.

\subsubsection{Geometry + dynamics}

A more realistic treatment can be obtained using the survival factor $P_{NH}(b)$ that describes the probability of no additional hadronic interaction between the nuclei, which is usually estimated using the Glauber formalism. In this way  we take into account the fact that, even when $b \gtrsim 2R$,  the probability of having strong interactions is finite. In this formalism, $S_{abs}^2(b)$ can be expressed in terms of the interaction probability between the nuclei at a given impact parameter, $P_H(b)$,  given by \cite{baltz_klein}

\begin{equation}
    S_{abs}^2(\mathbf{b}) \,=\, P_{NH}(\mathbf{b}) \,=\, 1 - P_H(\mathbf{b})
    \, ,
\label{eq:probability}
\end{equation}
where
\begin{equation}
    P_H(\mathbf{b}) = 1 - \exp
    \left[
        -\sigma_{nn} \int \mbox{d}^2\mathbf{r} \,
        T_A(\mathbf{r})T_A(\mathbf{r}-\mathbf{b})
    \right]
    \, ,
\label{eq:glauber}
\end{equation}
with $\sigma_{nn}$ being the total hadronic interaction cross section and $T_A$ the nuclear thickness function. As in Ref. \cite{baltz_klein}, we will assume that $\sigma_{nn} = 88$ mb at the LHC. It is interesting to observe that, since $\sigma_{nn}$ grows with the energy, the interaction probability also grows with the energy for a fixed impact parameter. In principle, this might lead to big differences between the results obtained with (\ref{eq:Sabs_baur}) and (\ref{eq:probability}). However,  $\sigma_{nn}$  grows with the energy as $\propto ln s$ or at most as $\propto ln^2 s$  and the differences between the two approaches turn out to be modest. 

As demonstrated in \cite{Azevedo:2019fyz}, the main difference between the absorption models is that the description of the absorptive factor given by Eq. (\ref{eq:probability}) implies a smooth transition between the small ($b<2R$) and large ($b>2R$) impact parameter behavior. For comparison purposes, in what follows we will also consider the case without absorption effects, where $S^2_{abs}(b) = 1$.  

\subsubsection{Kinematics} 

The absorptive factors (\ref{eq:Sabs_baur}) and (\ref{eq:probability}) emphasize the geometrical and dynamical aspects involved in the operational definition of an UPC. There is yet a third way to define an UPC in terms of a kinematical constraint, as proposed
in \cite{vy}. An ultra-peripheral collision can also be defined in terms of the momentum of the photons involved in the interaction. 
The distribution of equivalent photons generated by a moving particle with the charge $Ze$ is \cite{bertulani:1987tz, bertulani:2005ru, baltz:2007kq}:
\beq
N({\bf q})  = \frac{Z^2 \alpha}{\pi^2} \frac{({\bf q}_{\perp})^2}
{\omega \, q^4}  = \frac{Z^2 \alpha}{\pi^2 \omega}
\frac{({\bf q}_{\perp})^2}{\left( ({\bf q}_{\perp})^2
+ (\omega/\gamma)^2\right)^2}   \, ,
\label{defn}
\eeq
where  $q$ is the photon 4-momentum, ${\bf q}_{\perp}$ is its transverse
component, $\omega$ is the photon energy.  To obtain the equivalent photon spectrum, one has to integrate this 
expression over the transverse momentum up to some value $\hat{q}$.  After integrating  over the photon transverse momentum, the equivalent photon energy spectrum is given by:
\begin{equation}
    N(\omega) = 
    \frac{2Z^2\alpha}{\pi} \, 
    \ln 
    \left(
        \frac{\hat{q}\gamma}{\omega}
    \right) 
    \frac{1}{\omega}
    \, ,
\label{eq:flux_qchapeu}
\end{equation}
where the value of $\hat{q}$ must be chosen such that the particle emitting the photon does not break apart when emitting a photon with that momentum \cite{vy}.  In an UPC, photon emission is a coherent process, i.e., the photon is emitted by the whole source with a radius $R$. 
Therefore, the coherent photon wavelength is at least of order $R$ and we can interpret $\hat{q}$ as its maximum virtuality \cite{baur1998}. This gives us, in a first approximation, an estimate of $\hat{q} = \hbar c/ R$. For Pb, $R \approx 7$ fm, and hence
$\hat{q} \approx 0.028$ GeV. If $\hat{q}$ is larger than that, the photon starts to "resolve" the source and it might be emitted by a part of
the source. In order to have an idea of the sensitivity of the results to this choice, we will take $\hat{q}$ to be in the range $ 0.014 < \hat{q} < 0.028$ GeV.
These numbers are of the order of the binding energy of a nucleon in the nucleus. If $\hat{q}$ was larger, during the interaction the 
photon emission might induce the recoil of the emitting part of source (a nucleon) and cause its expulsion from the nucleus. In this case the
the final state would contain fragments and  would not be equal to the initial state, contradicting the definition of an UPC. These  arguments 
are very qualitative, but they set a scale for $\hat{q}$.  As shown in \cite{sabn}, the dependence of the results on the choice of $\hat{q}$ 
is very strong and hence the predictions should be given always with a corresponding uncertainty. On the other hand, the value of $\hat{q}$ depends only on the photon source and, in this sense, it is universal, i.e., the same for a wide variety of  photon-photon reactions with different final states. Therefore it can be determined studying measured reactions and then used to predict cross sections of yet unobserved processes. 

Since the photon flux  (\ref{eq:flux_qchapeu}) does not depend on the impact parameter, (\ref{eq:total_cross_section_etab}) simplifies to:
\begin{align}
	\sigma (Pb \, Pb \rightarrow Pb \, Pb \,  \eta_b) &= \int\limits_{m_{\eta}^2/\hat{q}\gamma}^
	{\hat{q}\gamma}d\omega_1\int\limits_{m_{\eta}^2/\omega_1}^{\hat{q}\gamma}
	d\omega_2 \, \hat{\sigma}_{\gamma \gamma \to \eta_b}(\omega_1,\omega_2) \, N(\omega_1) \, N(\omega_2),
	\label{sigmabs2}
\end{align}
where $m_{\eta}= m_{\eta_b} $.  

In principle, the three definitions mentioned above are related to each other. They all refer, in a way or another, to the size of the source. 
We could even say that pure geometry and kinematics are related through a Fourier transform. In practice however, these prescriptions are 
employed independently and it is useful to check whether they lead to equivalent results.

\section{Results and Conclusions}

In Fig. \ref{fig:energy_rapidity}a we present our predictions for the energy dependence of the total cross section for $\eta_b$ production in ultra-peripheral collisions $PbPb$, obtained using the geometric absorption factor (\ref{eq:Sabs_baur}) 
and different models for the form factor. 
For comparison, we also include the predictions for the production of $\eta_c$.

In Fig. \ref{fig:energy_rapidity}b we show the rapidity distribution of $\eta_b$ produced in $PbPb$ collisions at $\sqrt{s} = 5.02$ TeV. As
expected, the maximum of distribution occurs  at central rapidities, with the monopole (point-like) predicting larger (smaller) values in comparison to the more precise prediction derived using the realistic form factor. Moreover, the predictions for the $\eta_b$ meson are characterized by smaller normalizations and narrower rapidity distributions than those for the $\eta_c$ meson. 

\begin{figure}
    \centering
    \includegraphics[width=.60\linewidth]{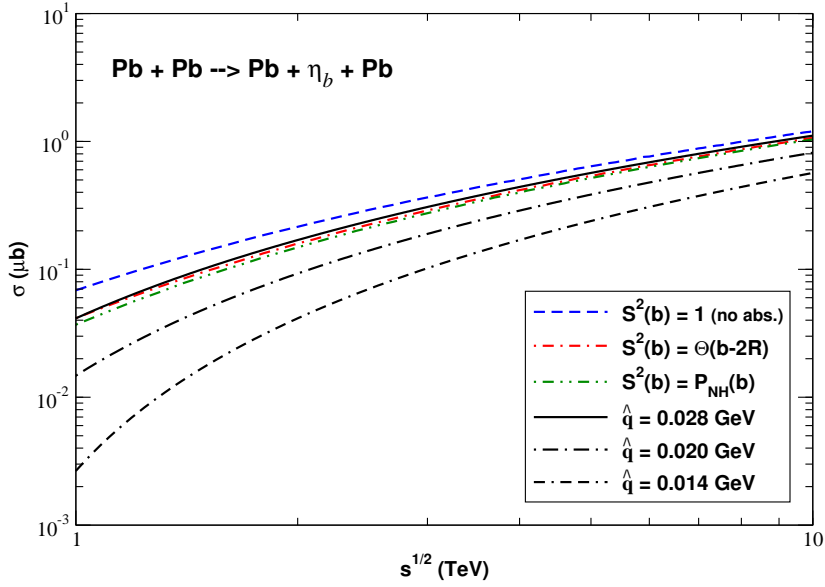}
    \caption{Energy dependence of the total cross section (in $\mu$b) for $\eta_b$ production in  $PbPb$ collisions calculated with (\ref{sigmabs2}) for three different values of $\hat{q}$. The blue, red and green  lines were obtained with (\ref{eq:total_cross_section_etab}) and  with the realistic form factor (\ref{eq:realistic}). }
    \label{fig:energy_qchapeu}
\end{figure}

In  Fig. \ref{fig:energy_qchapeu} we show the energy dependence of the total cross section for $\eta_b$ production in ultra-peripheral collisions $PbPb$ calculated with (\ref{sigmabs2}) for three different values of $\hat{q}$. We also show the result obtained with (\ref{eq:total_cross_section_etab}) and the realistic form factor (\ref{eq:realistic}) with the geometric absorption factor (\ref{eq:Sabs_baur}) and the survival factor $P_{NH}(b)$ (\ref{eq:probability}). For completeness, the prediction obtained disregarding the absorptive effects $(S^2_{abs}(b)=1)$ is also presented.

Having shown the energy dependence, let us focus on  $\sqrt{s} = 5.02$ TeV,  which is the most interesting case and where we may have data. 
In Table \ref{tab:results_cross_section} we show all the possible ways to compute the cross section with the different ingredients. 
In all cases $\Gamma(\eta_b \rightarrow \gamma\gamma) = 0.557 $ keV.  For comparison, we also show the results for $\eta_c$. 
As expected, the upper limit of the cross sections is set by $S^2_{abs}(b)=1$, i.e., when there is no condition  
enforcing  the ultra-peripheral nature of the collision. For all  absorption factors, the pointlike and monopole form factors set, respectively, the lower and upper limit of the cross section. The results obtained with the realistic form factor  (\ref{eq:realistic})
lie in between. Since the predictions obtained with the geometric and geometric+dynamic absorption factor nearly coincide, we can say 
that the most likely value of the cross section is $0.52 - 0.54$ microbarns. The error can be estimated from the Table, since the different choices of the form factors give a good representation of the uncertainty. These numbers are compatible with those found using the 
$\hat{q}$ prescription. The latter present a larger uncertainty. Taken together, these numbers show a nice convergence to the 
value of the cross section mentioned above, which is in principle large enough to be observed. 

In Table \ref{tab:lit_cross_section} we present a compilation of previous estimates of the same quantities. 
These cross sections have been calculated by several authors. In particular, the  formalism used  in Refs. \cite{baur1990} and \cite{chikin2000} is quite simlar to the one used here. Comparing the results shown in Table \ref{tab:results_cross_section} and Table \ref{tab:lit_cross_section} we see that they are consistent.  The least known ingredient was $\Gamma(\eta_b \rightarrow \gamma\gamma) $ 
and in the quoted previous works it was assumed to be $\simeq 0.41$ keV. The lattice QCD  value is $\simeq 0.55$ keV, i.e. ~ 34 \% larger 
and  approximately so are the corresponding cross sections. 

To summarize, we have updated the estimate of the $\eta_b$ production cross section in  ultra-peripheral Pb Pb collisions at $\sqrt{s} = 5.02$ TeV per nucleon pair.  The two main improvements were the introduction of a very recently calculated value of 
$\Gamma(\eta_b \rightarrow \gamma\gamma) $  and also the systematic comparison of different prescriptions to treat the absorptive effects.
The obtained values of the $\eta_b$ production cross section  are larger than the previous ones but within the same order of magnitude.  
Finally, it is reassuring to observe the three central lines in Fig. \ref{fig:energy_qchapeu} and realize that the three ways to define an 
UPC lead to very similar results.

\begin{figure}[t]
\begin{tabular}{cc}
\includegraphics[width=.47\linewidth]{./figs/energia_etab_etac.eps} \;& 
\includegraphics[width=.47\linewidth]{./figs/rapidez_etab_etac_theta.eps} \;\;\;\;\; \\
  (a) & (b)
\end{tabular}
\caption{(a) Energy dependence of the total cross section and (b) rapidity distribution for $\eta_b$ photoproduction in ultra-peripheral $Pb Pb$ collisions. The predictions for the $\eta_c$ in the final state are also presented (red lines) for comparison.}
\label{fig:energy_rapidity}
\end{figure}


\begin{table}
\begin{center}
\renewcommand{\arraystretch}{1.4} 
\begin{tabular}{|| c | c | c  c  c ||}
\hline 
\hline 
\multirow{2}{*}{\bf{Meson}} &  
\multirow{2}{*}{\bf{Absorptive factor}}  & \multicolumn{3}{c||}{\multirow{2}{*}{$\;\sigma \; (\mu \mbox{b})\;$}} \\ 
 & & & & \\
\hline
\hline
 & & $\quad\quad \mbox{point-like} \quad$ & $\quad \mbox{monopole} \quad$ & $\quad \mbox{realistic} \quad\quad$ \\ 
\cline{3-5}
                & $\; S^2_{abs}(\textit{\textbf{b}})=1 \rightarrow \mbox{no abs.} \;$ & 0.43 & 0.76 & 0.64 \\
\multirow{2}{*}{$\quad \eta_b (1S) \quad$}   & $S^2_{abs}(\textit{\textbf{b}})=\Theta (|\textit{\textbf{b}}|-2R)$ & 0.39 & 0.62 & 0.54 \\
                & $S^2_{abs}(\textit{\textbf{b}}) = 1 - P_H(\textit{\textbf{b}}) \;\;$ & 0.38 & 0.59 & 0.52 \\
\cline{2-5}
& \multirow{2}{*}{$\hat{q}$} & $\quad\quad \hat{q} = 0.014\;\mbox{GeV} \quad$ & $\quad \hat{q} = 0.020\;\mbox{GeV} \quad$ & $\quad \hat{q} = 0.028 \;\mbox{GeV} \quad\quad$ \\
\cline{3-5}
& & 0.24 & 0.39 & 0.57\\
\hline\hline
& & $\quad\quad \mbox{point-like} \quad$ & $\quad \mbox{monopole} \quad$ & $\quad \mbox{realistic} \quad\quad$ \\ 
\cline{3-5}
                        & $\; S^2_{abs}(\textit{\textbf{b}})=1 \rightarrow \mbox{no abs.} \;$ & 368.62 & 557.39 & 497.03 \\
\multirow{2}{*}{$\quad \eta_c (1S) \quad$}    & $S^2_{abs}(\textit{\textbf{b}})=\Theta (|\textit{\textbf{b}}|-2R)$ & 348.66 & 495.25 & 451.22 \\
                        & $S^2_{abs}(\textit{\textbf{b}})=1 - P_H(\textit{\textbf{b}}) \;\;$ & 343.72 & 485.04 & 441.94 \\
\cline{2-5}
& \multirow{2}{*}{$\hat{q}$} & $\quad\quad \hat{q} = 0.014\;\mbox{GeV} \quad$ & $\quad \hat{q} = 0.020\;\mbox{GeV} \quad$ & $\quad \hat{q} = 0.028 \;\mbox{GeV} \quad\quad$ \\
\cline{3-5}
& & 259.65 & 355.74 & 465.77 \\
\hline
\hline
\end{tabular}
\caption{Total cross sections (in $\mu$b) for  $\eta_b$ and $\eta_c$ production in $\gamma \gamma$ interactions in ultra-peripheral 
$PbPb$ collisions at $\sqrt{s} = 5.02$ TeV.}
\label{tab:results_cross_section}
\end{center}
\end{table}

\begin{table}[h!]
\centering
\renewcommand{\arraystretch}{1.4} 
\begin{tabular}{|c|c|c|c|c|}
\hline
\hline
\multirow{2}{*}{$\;\mbox{Meson}\;$} &  
\multirow{2}{*}{$\; \Gamma_{\gamma\gamma} \; \mbox{[KeV]} \;$}  & \multicolumn{3}{c|}{$\;\sigma \; (\mu \mbox{b})\;$} \\ \cline{3-5}
 & &  $\; Z=82 \quad \gamma = 2750 \;$ &  $\; Z=82 \quad \gamma = 3000 \;$ & $\; Z=82 \quad \gamma = 4000 \;$ \\ 
\hline
\hline
                       & \multirow{2}{*}{0.40} & 0.322 \cite{chikin2000}  &  0.445 \cite{vidovic1995} & \\
                       &  & 0.466 \cite{krauss1997}    & & \\
                       \cline{2-5}
\multirow{3}{*}{$\quad \eta_b (1S) \quad$} & \multirow{3}{*}{0.41} &                         & & 0.46 \cite{baur1990}  \\
                       & &                         & & 0.50 \cite{chikin2000} \\
                       & &                         & & 0.90 \cite{bertulani1989} \\
                       \cline{2-5}
                       & \multirow{2}{*}{0.43} & 0.346 \cite{chikin2000} &   & \\
                       &  & 0.37 \cite{baur1998} & & \\
\hline
\hline
                        & \multirow{2}{*}{6.30} & 464 \cite{chikin2000} & 603 \cite{vidovic1995} &590 \cite{baur1990}\\
                        &  & 598 \cite{krauss1997} &                   &644 \cite{chikin2000} \\
                        &  &                       &                   &800 \cite{bertulani1989} \\
                        \cline{2-5}
\multirow{-3}{*}{$\quad \eta_c (1S) \quad$} & \multirow{3}{*}{7.50} & 552 \cite{chikin2000} & & \\
                        &  & 590 \cite{baur1998}   & & \\
\hline
\hline
\end{tabular}
\caption{Total cross sections (in $\mu$b) for $\eta_b$ and $\eta_c$ production in $\gamma \gamma$ interactions obtained in previous works.}
\label{tab:lit_cross_section}
\end{table}

\begin{acknowledgments} 
This work was  partially financed by the Brazilian funding
agencies CNPq, CAPES, FAPESP and INCT-FNA (process number          
464898/2014-5). F.S.N.  gratefully acknowledges the  support from the 
Funda\c{c}\~ao de Amparo \`a  Pesquisa do Estado de S\~ao Paulo (FAPESP).

\end{acknowledgments}




\end{document}